\documentclass[final,3p]{CSP}

\usepackage{amsmath, amssymb, amsfonts, amsthm, latexsym, bm, mathdots, accents}
\usepackage{graphicx}
\usepackage{float}
\usepackage{tikz}
\usetikzlibrary{matrix,positioning,cd}
\usepackage{bbm}
\usepackage{dsfont}
\usepackage{braket}
\usepackage{yfonts}
\usepackage[mathscr]{euscript}
\usepackage{siunitx}
\usepackage{booktabs}
\usepackage{changepage}
\usepackage{subcaption}
\usepackage{color}
\usepackage{url}
\usepackage{breqn}

\usepackage[utf8]{inputenc}

\usepackage{hyperref}

\usepackage{caption}

\usepackage{dcolumn}

\begin{document}

\begin{frontmatter}

\title{Evolution of Hawking mass under perturbative spacetime uniformly expanding flows }

\author[mymainaddress,currentaddress]{Hollis Williams\corref{cor1}}

\address[mymainaddress]{%
Theoretical Sciences Visiting Program, 
Okinawa Institute of Science and Technology Graduate University \\ 
Onna-son, Okinawa 904-0495, Japan
}

\address[currentaddress]{%
Department of Mathematics and Statistics, University of Exeter, Exeter EX4 4QF, UK
}

\cortext[cor1]{Email: holliswilliams@hotmail.co.uk}

\begin{abstract}\rm
\begin{adjustwidth}{2cm}{2cm}{\itshape\textbf{Abstract:}} We present a numerical investigation of the evolution of the Hawking mass for perturbed surfaces evolving under hypersurface-restricted uniformly expanding flows in Minkowski spacetime.  Although monotonicity of the Hawking mass under inverse mean curvature flow is well understood, much less is known about the behaviour of such flows in genuine spacetime contexts.  To move beyond the totally geodesic setting where uniformly expanding flows reduce to Euclidean inverse mean curvature flow, we introduce controlled perturbations of the ambient extrinsic curvature. This yields a hypersurface-restricted realization of a spacetime uniformly expanding flow with a nontrivial mean curvature vector. Our results indicate that monotonicity of the Hawking mass remains stable under a range of perturbation amplitudes, angular modes, and spacetime deformations.  These results provide evidence for robustness of monotonicity and establish a computational framework for future investigations of uniformly expanding flows in more general spacetime geometries.
\end{adjustwidth}
\end{abstract}
\end{frontmatter}

\section{Introduction}

\noindent
Quasi-local notions of mass in general relativity play a central role in understanding the distribution of energy in spacetimes, since they provide a way to associate an energy content to a spacelike 2-surface using its intrinsic and extrinsic geometry \cite{dougan, szabados2, brown, yau, szabados, murchadha, liu}.  Amongst the various proposed definitions, the Hawking mass has attracted interest due to its simple geometric definition and its close connection to the Penrose inequality and gravitational collapse \cite{hawking, penrose, mars}.  Although the Hawking mass can be negative in general, recent work has shown that it is non-negative and possesses a rigidity property when evaluated on suitable area-constrained critical surfaces if the dominant energy condition is also assumed \cite{diaz}.  A major advantage of the Hawking mass is its desirable properties under geometric flows: formal monotonicity of the mass under inverse mean curvature flow (IMCF) was a key ingredient in the first proof of the Riemannian Penrose inequality \cite{huisken, bray}.  Uniformly expanding flows (UEFs) extend this idea to spacetime by prescribing an evolution of surfaces with constant expansion, allowing both spacelike and timelike components of the mean curvature vector to contribute \cite{hayward1, frau, malec, hayward2}. In this framework, the notion of a time-flat surface was introduced and variational formulas have been derived showing that the Hawking mass is formally nondecreasing along spacelike UEFs under suitable conditions \cite{jeffrey, jeffrey2}.  These flows are a natural way to investigate monotonicity properties of quasi-local masses and may be considered as a generalization of the IMCF. 

However, these monotonicity results are proved under the assumption that a smooth flow exists: no general existence theory is known, and no analysis has been performed for surfaces deviating from perfect symmetry, with less known about the stability of this behaviour. In particular, no numerical or perturbative studies have tested the robustness of monotonicity under non-spherical deformations.  On the formal side, a full existence theory for general spacelike UEFs remains challenging due to the forward-backward parabolic nature of the associated system of PDEs.  In this work, we take a complementary computational approach.  Rather than attempting to construct the most general uniformly expanding flow in a spacetime, we focus on a physically natural and mathematically well-defined subclass: closed spacelike 2-surfaces embedded in Minkowski spacetime.  Our framework proceeds in two stages.  We first study the time-flat setting, where the flow reduces to IMCF and provides a convenient validation regime.  We then introduce controlled spacetime perturbations through a small addition to the ambient extrinsic curvature.  In the totally geodesic setting, the spacetime mean curvature vector reduces back to the hypersurface normal.  This allows us to evolve a perturbed surface numerically under an expanding flow and track the evolution of the Hawking mass, providing evidence that monotonicity persists beyond an idealized setting.

The simplest and most natural spacetime which we may consider in this setting is a perturbation of a spherical surface in Minkowski space.  Minkowski space appears as the unique zero-mass configuration in general relativity with perturbations necessarily increasing the total energy, a fact which underlies the original spinor-based proof of the positive energy theorem \cite{witten, taubes}.  Proofs of this theorem first appeared for small perturbations of Minkowski spacetime, clarifying the role that flat spacetime should play in the general result \cite{geroch, yvonne}.  For quasi-local masses, we might also expect failures of positivity or monotonicity to first manifest for non-spherical perturbations of flat space.  Consequently, verifying monotonicity for small deviations from spherical symmetry in Minkowski spacetime constitutes an essential consistency check for UEFs and their suitability as physical diagnostics.  We therefore expect that the behavior of the Hawking mass under perturbations of round spheres in Minkowski space provides a sensitive probe of both rigidity and stability.

Using spherical harmonics to parametrize perturbations of a sphere evaluated on a finite $(\theta,\phi)$ grid, we calculate the Hawking mass along the flow and examine its sensitivity to perturbations of the spacetime and the angular structure of the harmonics.  The use of spherical harmonic decomposition is standard in perturbative and numerical relativity, where they are typically used to achieve spectral convergence and isolate angular modes \cite{novak,bishop}.  Representations based on spherical harmonics are also widely used in numerical studies of gravitational perturbations and wave extraction \cite{kokkotas, motl2, motl1, chung}.  Restricting to low-order harmonic modes allows us to probe controlled deviations from spherical symmetry whilst retaining a clear interpretation of the resulting geometric and physical quantities.  Moreover, expressing perturbations in an orthonormal harmonic basis enables a systematic assessment of how individual angular modes influence the evolution of the Hawking mass, providing a framework in which to test the robustness of monotonicity under nonspherical deformations.

We begin by validating the numerical framework in the time-flat setting, where the flow reduces to an in-slice IMCF.  We then introduce controlled spacetime perturbations through an addition to the ambient extrinsic curvature, moving beyond the totally geodesic setting.  The resulting evolution should therefore be viewed as a hypersurface-restricted spacetime expanding flow rather than a purely Euclidean geometric flow.  In the numerical implementation, the evolution equation is realized through a scalar graph formulation where the graph factor is partly absorbed into the definition of the approximate mean curvature. The resulting scheme should therefore be viewed as a model of a spacetime expanding flow rather than a complete discretization of the full UEF equations, which is beyond the scope of this work.  In contrast to the time-flat setting where $K_{ij}=0$, our model includes a contribution to the extrinsic curvature via an effective expansion scalar.  The resulting evolution therefore probes Hawking mass monotonicity in a genuine spacetime setting, although the full UEF system is not imposed.  The numerical results presented here provide evidence that Hawking mass monotonicity remains stable under a class of nonspherical perturbations in both the time-flat and spacetime-perturbed settings considered.  In particular, the observed behaviour suggests that the monotonicity properties associated with UEFs are not artifacts of spherical symmetry and persist under controlled deviations.

The article is structured as follows.  In Section 2, we provide background and definitions for the Hawking mass, UEFs, time-flat surfaces, in-slice flows, and the scalar graph formulation of the flow.  In Section 3, we outline the numerical framework used throughout the paper.  Section 4 presents a series of validation tests, including convergence studies and comparisons with known analytic results for round and perturbed spheres in Minkowski spacetime.  In Section 5, we present numerical results for spacetime-perturbed expanding flows, examining the evolution and monotonicity of the Hawking mass under variations in perturbation amplitude, angular structure, and ambient extrinsic curvature.  We finish with conclusions and future directions in Section 6.

\section{Background}

\subsection{Hawking mass}

\noindent
Let $\Sigma$ be a smooth, closed, spacelike two-surface embedded in a four-dimensional spacetime $(\mathcal{M},g)$. Denote by $|\Sigma| $ the area of $\Sigma$, and by $\vec{H}$ its mean curvature vector, defined as the trace of the second fundamental form with respect to the induced metric on $\Sigma$.  The Hawking mass $m_H$ is defined to be

\begin{equation}
m_H(\Sigma)
=
\sqrt{\frac{|\Sigma| }{16\pi}}
\left(
1 - \frac{1}{16\pi}
\int_\Sigma \braket{\vec{H}, \vec{H}}  \, dA
\right),
\end{equation}

\noindent
where $dA$ is the area element induced on $\Sigma$.  In this work we restrict to closed orientable surfaces with spherical topology, for which the standard normalization in the definition of the Hawking mass applies.  It can be checked with standard calculations that $m_H =0$ in the case of a round sphere in Minkowski spacetime and that $m_H = m$ for a round sphere in Schwarzschild spacetime with mass $m$.  To explore the behavior of the Hawking mass under small deformations of a spherically symmetric surface, we consider perturbations of a round sphere in the chosen time-symmetric slice. We parametrize the perturbed surface as a radial graph over the sphere:

\begin{equation}   
r ( \theta, \phi) = R_0  [1 + \epsilon Y_{lm} ( \theta, \phi)], \:\:\:\:\:\:   \epsilon << 1  ,       \end{equation}

\noindent
where $R_0$ is the radius of the sphere, $\epsilon$ is a small dimensionless parameter, and $Y_{lm} ( \theta, \phi)$ are spherical harmonics.  This representation allows us to study the perturbation analytically mode by mode and is also the starting point for the analytic calculation of the Hawking mass $m_H$ at time $t=0$ whose evolution under the flow will be studied numerically.

To calculate the Hawking mass at $t=0$ for a perturbed sphere in Minkowski spacetime $ m^M_H(t=0)$, one may work in the constant $t$ slice of Minkowski with spherical coordinates $(r, \theta, \phi)$ and expand to first order in $\epsilon$.  For a radial graph, the induced metric $g_{AB}$ is

\begin{equation} g_{AB} = r^2 \Omega_{AB} + r_A r_B ,   \end{equation}

\noindent
where $\Omega_{AB}$ is the unit sphere metric and $r_A = \partial_A r$.  Since the integral of $Y_{lm}$ over the two-sphere $S^2$ vanishes for $l \geq 1$, we have

\begin{equation} d A = r^2 \sin \theta \: d \theta \: d \phi + \mathcal{O} ( \epsilon^2 ),   \end{equation}

\begin{equation}  |\Sigma| =  4 \pi R_0^2 + \mathcal{O} ( \epsilon^2 )           . \end{equation}

\noindent
Using the expansion

\begin{equation}  H = \frac{2}{r} - \frac{1}{r^2} \Delta_{S^2} r + \mathcal{O} ( | \nabla r |^2 )  ,         \end{equation}

\noindent
we then have

\begin{equation} H = \frac{2}{R_0} + \epsilon \bigg( \frac{l (l+1) -2}{R_0}  \bigg) Y ( \theta, \phi) + \mathcal{O} (\epsilon^2 ) ,    \end{equation}

\noindent
hence the mean curvature has a linear correction which is proportional to the spherical harmonics.  Carrying out the surface integral, we then arrive at 

\begin{equation}  m_H^M (t=0 ) = \mathcal{O} ( \epsilon^2 ) .       \end{equation}

%\[  m_H^M (t=0 ) =  \frac{R_0}{2} \epsilon^2  + \mathcal{O} ( \epsilon^4 ) .       \]

Performing the analogous calculations for a perturbed sphere in Schwarzschild spacetime with the usual time-symmetric slice

\begin{equation}
     d s^2   = \bigg( 1 - \frac{2m}{r} \bigg)^{-1} dr^2 + r^2 d \Omega^2 ,    \end{equation}

\noindent
we one can also check as expected that

\begin{equation} m_H^S (t = 0) = m + \mathcal{O} ( \epsilon^2)  . \end{equation}

\noindent
This result for the corrected Hawking mass provides an analytic baseline for comparison with numerical calculations of its evolution under the chosen flow.

\subsection{Uniformly expanding flows}

\noindent
Assume that $\vec{H}$ is spacelike, so that its magnitude
\begin{equation}
|\vec{H}| = \sqrt{\braket{\vec{H}, \vec{H}} }
\end{equation}
is well defined.  A one-parameter family of surfaces $\{\Sigma_\lambda\}$ is said to evolve by a UEF if the flow vector field $\vec{\xi}$ normal to $\Sigma_\lambda$ satisfies
\begin{equation}
\vec{\xi} = \frac{\vec{H}}{|\vec{H}|^2},
\end{equation}

\noindent
or, more generally, if the expansion of the surface along $\vec{\xi}$ is constant on each leaf of the flow. In the Riemannian setting, this reduces to IMCF, where the normal speed is given by the reciprocal of the mean curvature scalar.  In spacetime, UEFs generalize IMCF by allowing both spacelike and timelike components of the normal evolution whilst preserving uniform control of the surface expansion.

\subsection{Time-flat surfaces}

\noindent
Let $\{ \nu_1, \nu_2 \}$ be a local orthonormal frame of the normal bundle $N\Sigma$, with $\nu_1$ chosen to be proportional to the mean curvature vector $\vec{H}$. The normal connection one-form $\alpha_H$ associated with $\vec{H}$ is defined by
\begin{equation}
\alpha_H(X) = \braket{ \nabla_X \nu_1 , \nu_2 },
\qquad X \in T\Sigma.
\end{equation}

\noindent
A spacelike two-surface $\Sigma$ is said to be time-flat if
\begin{equation}
\text{div} \:\alpha_H = 0.
\end{equation}

\noindent
Geometrically, this condition implies that the mean curvature vector does not rotate within the normal bundle and that the normal evolution of the surface does not twist. This is the condition under which additional normal bundle contributions to the evolution of the Hawking mass vanish, motivating its use in numerical investigations of monotonicity.  In the present work, we do not attempt to solve the full time-flat constraint system directly.  We instead employ a scalar graph realization of the flow which reproduces the time-flat IMCF regime in the absence of spacetime perturbations.

\subsection{In-slice  flows}

\noindent
 Let $H$ denote the mean curvature scalar of $\Sigma$ computed with respect to $h$, and let $\nu$ be the outward unit normal to $\Sigma$ within $M$.  An in-slice IMCF is defined by the evolution equation
\begin{equation}
\frac{\partial x}{\partial \lambda} = \frac{1}{H} \, \nu,
\end{equation}
where $x(\lambda)$ denotes the embedding of the evolving surface $\Sigma_\lambda$ into $M$ and $\lambda$ is a flow parameter labeling the leaves of the evolution.  Although the evolution is confined to a fixed spacelike hypersurface, the Hawking mass is computed using the spacetime mean curvature vector.

\subsection{Radial graph formulation and scalar evolution equation}

\noindent
To implement the in-slice IMCF numerically, we represent the evolving
surface as a radial graph over the unit sphere
\begin{equation}
X(\theta,\phi,\lambda)
=
r(\theta,\phi,\lambda)\,\omega(\theta,\phi),
\end{equation}
where $\omega(\theta,\phi)$ denotes the outward unit radial vector on
$S^2$ and $\lambda$ is the flow parameter.  Tangent vectors to the
surface are then defined as
\begin{equation}
X_A
=
\partial_A X
=
r_A \omega + r\,\omega_A,
\end{equation}
where $A \in \{\theta,\phi\}$ and $r_A = \partial_A r$.
Using the standard relations
\begin{equation}
\langle \omega,\omega\rangle = 1,
\qquad
\langle \omega,\omega_A\rangle = 0,
\end{equation}
the induced metric on the surface takes the form
\begin{equation}
g_{AB}
=
r_A r_B + r^2 \sigma_{AB},
\end{equation}
where $\sigma_{AB}$ is the metric on the unit sphere.

The outward unit normal vector field $\nu$ is determined by the condition
\begin{equation}
\langle \nu , X_A \rangle = 0.
\end{equation}
In this work, we restrict to star-shaped surfaces.  This guarantees that each outward radial ray intersects the evolving surface exactly once and allows the flow to be expressed as a scalar evolution equation for the graph function $r(\theta,\phi,\lambda)$.  For such a star-shaped radial graph one obtains
\begin{equation}
\nu
=
\frac{1}{W}
\left(
\omega
-
\frac{1}{r}\nabla_{S^2} r
\right),
\end{equation}
where
\begin{equation}
W
=
\sqrt{
1+\frac{|\nabla_{S^2} r|^2}{r^2}
}.
\end{equation}

\noindent
The in-slice IMCF equation
\begin{equation}
\frac{\partial X}{\partial \lambda}
=
\frac{1}{H}\,\nu
\end{equation}
therefore induces a scalar evolution equation for the graph function
$r(\theta,\phi,\lambda)$.  Since
\begin{equation}
\frac{\partial X}{\partial \lambda}
=
\frac{\partial r}{\partial \lambda}\,\omega,
\end{equation}
taking the inner product with $\omega$ gives
\begin{equation}
\frac{\partial r}{\partial \lambda}
=
\frac{1}{H}
\langle \nu,\omega\rangle.
\end{equation}
Using
\begin{equation}
\langle \nu,\omega\rangle
=
\frac{1}{W},
\end{equation}
we arrive at the scalar graph evolution equation
\begin{equation}
\frac{\partial r}{\partial \lambda}
=
\frac{1}{H\,W}
\end{equation}
with
\begin{equation}
W
=
\sqrt{
1+\frac{|\nabla_{S^2} r|^2}{r^2}
}.
\end{equation}

\noindent
For sufficiently small perturbations of a round sphere,
$|\nabla_{S^2} r| \ll r$, one has $W = 1 + \mathcal{O}(|\nabla r|^2)$,
so the evolution reduces perturbatively to the simpler form
$\partial_\lambda r \approx 1/H$.  In the numerical framework outlined in the next section, the evolution equation is implemented via a scalar graph formulation where the graph factor $W$ is partly absorbed into the approximate mean curvature used in the discretization.  The resulting numerical scheme should therefore be viewed as a geometric approximation to the evolution equation rather than a  discretization of the full UEF system, which we leave for future work.

\subsection{Extension beyond the totally geodesic case}

\noindent
The preceding discussion applie to the totally geodesic setting, where the spacetime mean curvature vector reduces to the spatial mean curvature
normal within the hypersurface and the UEF reduces to an IMCF.  To probe genuine spacetime effects, we must also consider hypersurfaces with nonvanishing extrinsic curvature $K_{ij}$.  For a closed surface
$\Sigma$, the spacetime mean curvature vector may then be written
as
\begin{equation}
\vec H = H \nu - P n,
\end{equation}
where $H$ denotes the spatial mean curvature scalar of $\Sigma$ within $M$,
$\nu$ is the outward unit normal tangent to $M$, $n$ is the future-directed
timelike unit normal to $M$, and
\begin{equation}
P = \mathrm{tr}_{\Sigma} K
\end{equation}
is the trace of the extrinsic curvature over the tangent space of $\Sigma$.  The Lorentzian norm of the spacetime mean curvature vector is
\begin{equation}
|\vec H|^2 = H^2 - P^2.
\end{equation}
The Hawking mass therefore takes the form
\begin{equation}
m_H(\Sigma)
=
\sqrt{\frac{|\Sigma|}{16\pi}}
\left(
1-
\frac{1}{16\pi}
\int_\Sigma (H^2-P^2)\, dA
\right).
\end{equation}

\section{Numerical framework}

\noindent
As explained in Section 2.5, each evolving surface is represented as a
radial graph over the unit sphere.  In the time-flat setting, the evolution is governed by
the scalar graph equation
\begin{equation}
\frac{\partial r}{\partial \lambda}
=
\frac{1}{H\,W},
\end{equation}
with
\begin{equation}
W
=
\sqrt{
1+\frac{|\nabla_{S^2}r|^2}{r^2}
}.
\end{equation}

\noindent
To probe beyond the totally geodesic case,
we additionally include a nontrivial spacetime contribution via the
trace of the extrinsic curvature 
\begin{equation}
P = \mathrm{tr}_{\Sigma} K.
\end{equation}
The effective expansion scalar entering the evolution is then taken to be
\begin{equation}
H_{\mathrm{eff}}
=
H + P,
\end{equation}
yielding a modified flow equation
\begin{equation}
\frac{\partial r}{\partial \lambda}
=
\frac{W}{H_{\mathrm{eff}}}.
\end{equation}

\noindent
The quantity $P$ was modelled numerically as a perturbation of
the extrinsic curvature,
allowing the evolution to depart from in-slice IMCF whilst remaining
in a stable perturbative regime.  This provides a simplified
hypersurface-restricted implementation of a spacetime UEF.

Although perturbative expansions were discussed in Section 2 for analytic
comparison and validation purposes, the numerical implementation employs
nonlinear geometric quantities associated with the radial graph.  In particular, the induced metric, outward unit normal, area element, and mean curvature were computed directly from the embedding
$X(\theta,\phi,\lambda)$ without performing a linearization in the perturbation
amplitude.  The flow was integrated using an explicit Euler scheme with fixed time
step $\Delta \lambda$ chosen to be sufficiently small to ensure numerical
stability. Convergence of the evolution was verified by varying both the
time step and angular resolution.  The evolution was carried out over a
finite interval of the flow parameter sufficient for structure
to evolve under the flow operator and for numerical
instabilities to become apparent if present.  In selected runs, a small diffusion term was included to suppress
high-frequency numerical noise
\begin{equation}
\frac{\partial r}{\partial \lambda}
=
\frac{W}{H_{\mathrm{eff}}}
+
\eta\,\Delta_{S^2}r,
\end{equation}
where $\eta \ll 1$ is a regularization parameter.  Results were verified to be insensitive to small variations of $\eta$
within the regime considered.

Angular dependence on the surface was discretized using a uniform grid
in $(\theta,\phi)$, with $\theta$ taking values between $0$ and $\pi$ and $\phi$ taking value between $0$ and $2 \pi$.  Angular derivatives were computed using second-order finite-difference
approximations in both angular directions, yielding discrete
representations of the spherical gradient and Laplacian operators.  Spherical harmonics were used to specify the initial perturbations of
the surface geometry but the numerical evolution itself was performed directly on the
$(\theta,\phi)$ grid using finite-difference approximations rather than
a spectral harmonic decomposition.  The angular resolution was determined by the number of grid points
$(N_\theta,N_\phi)$. Numerical convergence was assessed by varying these
parameters and monitoring stability of the Hawking mass and related
geometric quantities for both round and perturbed surfaces.  Throughout the evolution we monitored:
\begin{itemize}
\item the surface area $|\Sigma_\lambda|$,
\item the integral $\int_{\Sigma_\lambda} (H^2-P^2)\, dA$,
\item the Hawking mass $m_H(\lambda)$.
\end{itemize}

\section{Validation}

\subsection{Round sphere in Minkowski spacetime}

\noindent
We first consider a round sphere of radius $R$ embedded in a spacelike hypersurface in Minkowski spacetime.
For this configuration, the Hawking mass vanishes identically.  The numerical implementation reproduces the exact area, mean curvature, and Hawking mass up to discretization error.  The results for the relevant geometric quantities are shown in Table 1.   As a further consistency check, we computed $m_H$ for varying mesh fineness, where $N_{\theta}$ and $N_{\phi}$ are the number of nodes in the $\theta$ and $\phi$ directions, respectively.  The computed value converges to zero with second-order accuracy under angular mesh refinement (shown in Table 2).  The computed Hawking mass remains small under angular mesh refinement,
with residual values reaching the $10^{-5}$ level at the higher working
resolutions.  The remaining deviation from zero is attributed to
finite-difference truncation and quadrature error associated with the
spherical discretization.  The small variations observed
between resolutions are consistent with the expected numerical
error once the dominant discretization effects become comparable
to machine and quadrature accuracy.

\begin{table}[h]
\centering
\begin{tabular}{lcc}
\hline
Quantity & Computed Value & Expected Value \\
\hline
$|\Sigma|$ & $1.256668\times10^{1}$ & $1.256637\times10^{1}$ \\
$H$ & $2.000000\times10^{0}$ & $2.000000\times10^{0}$ \\
$\int H^2\,\mathrm{d}A$ & $5.026670\times10^{1}$ & $5.026548\times10^{1}$ \\
$m_H$ & $1.215988\times10^{-5}$ & $0$ \\
\hline
\end{tabular}
\caption{Comparison of computed values against expected values for a round sphere in Minkowski spacetime.}
\label{tab:results_comparison}
\end{table}

\begin{table}[h]
\centering
\begin{tabular}{lcc}
\hline
$N_{\theta}$ & $N_{\phi}$ & $m_H$ \\
\hline
32 & 64 & $1.957659\times10^{-4}$ \\
64 & 128 & $ 3.424537\times10^{-4}$  \\
80 & 160 & $ -8.728824\times10^{-6}$  \\
96 & 192 & $-1.215988\times10^{-5}$ \\
\hline
\end{tabular}
\caption{Value for Hawking mass $m_H$ for differing values of $N_{\theta}$ and $N_{\phi}$.}
\label{tab:results_comparison}
\end{table}

In addition to validating these geometric quantities for static round spheres, we also evolved a round sphere in Minkowski spacetime under the in-slice IMCF implementation.  A round sphere in Minkowski spacetime has vanishing Hawking mass and under IMCF spherical symmetry is preserved throughout the evolution, hence the Hawking mass is expected to remain constant along the flow.  Numerically, we initialized the evolution with a round sphere of radius $R_0$ and monitored the Hawking mass over a finite interval of the flow parameter $\lambda$. Throughout the evolution, the computed Hawking mass remained close to its initial value, with only small fluctuations at the level of the discretization error and no systematic decrease in the Hawking mass. In particular, the discrete monotonicity diagnostic

\begin{equation}
\delta m_H
=
\max_n \min\!\left(m_H(\lambda_{n+1}) - m_H(\lambda_n),\, 0\right),
\end{equation}

\noindent
remained nonnegative within numerical tolerance (shown in Fig. 1). The observed stability confirms that the corrected scalar graph evolution equation and associated discretization reproduce the expected IMCF behaviour in the simplest symmetric setting. It also establishes a  baseline against which nonspherical and spacetime-deformed evolutions may subsequently be compared.

\begin{figure}
\centering
\includegraphics[width=100mm]{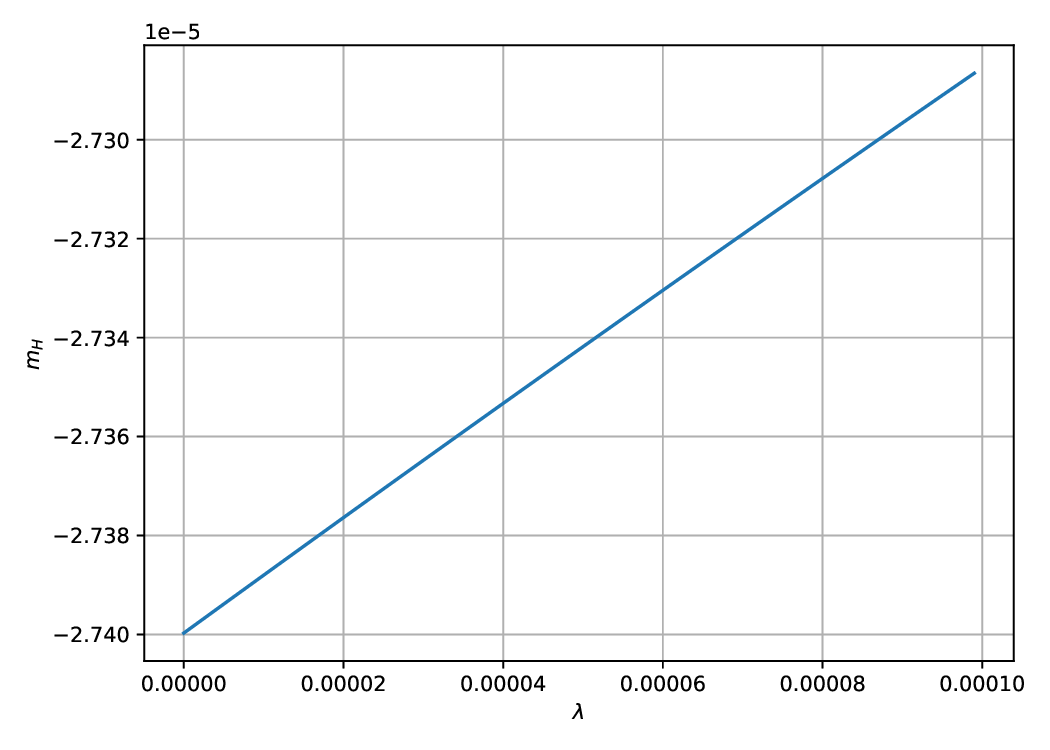}% 
\caption{\label{fig:epsart} Evolution of the Hawking mass for a round sphere in Minkowski spacetime under the corrected in-slice IMCF.  The numerical evolution preserves constant Hawking mass up to discretization error, with no systematic decrease observed throughout the evolution interval. }
\end{figure}

\subsection{Perturbed sphere in Minkowski spacetime}

\noindent
We next consider a small radial perturbation of the form
\begin{equation}
r(\theta,\phi) = R \left( 1 + \varepsilon Y_{\ell m}(\theta,\phi) \right)
\end{equation}
and compute the Hawking mass without evolving the surface.  We consider a single nontrivial spherical harmonic perturbation, choosing the axisymmetric mode $Y_{20}$. This provides a simple and representative deformation away from spherical symmetry but avoids unnecessary mode coupling.  For sufficiently small $\varepsilon$, the Hawking mass is expected to scale as $O(\varepsilon^2)$.  This behavior is confirmed numerically, as shown in Fig. 2.  The negative sign arises because small non-spherical perturbations of a round sphere in flat spacetime can lead to negative values of the Hawking mass.

Finally, we validate the scalar graph implementation of the in-slice IMCF by evolving a weakly perturbed sphere in Minkowski spacetime.  The initial surface was taken to be as above, with $\epsilon = 0.02$.  The evolution was performed using the scalar graph flow equation derived in Section 2 together with a small artificial viscosity term used to suppress high-frequency noise.  Unlike the round sphere case, the perturbed surface possesses nontrivial angular structure and therefore provides a more stringent test of the numerical implementation.  In particular, monotonicity of the Hawking mass is no longer guaranteed numerically by exact symmetry and instead depends on the stability and consistency of the flow.  The Hawking mass was monitored throughout the evolution interval together with the monotonicity diagnostic given by equation (39).  For the perturbed initial data considered here, the Hawking mass increased throughout the evolution up to numerical tolerance, with the largest decrease between successive time steps remaining at the level of $10^{-9}$.  No systematic violation or numerical instability was observed over the interval (as shown in Fig. 3).  These results provide evidence that the scalar graph formulation reproduces the expected qualitative behaviour of IMCF when exact symmetry is no longer imposed.  In particular, they demonstrate that the numerical framework preserves the monotonicity properties of the Hawking mass for small nonspherical perturbations, at least in the perturbative regime considered here.

\begin{figure}
\centering
\includegraphics[width=100mm]{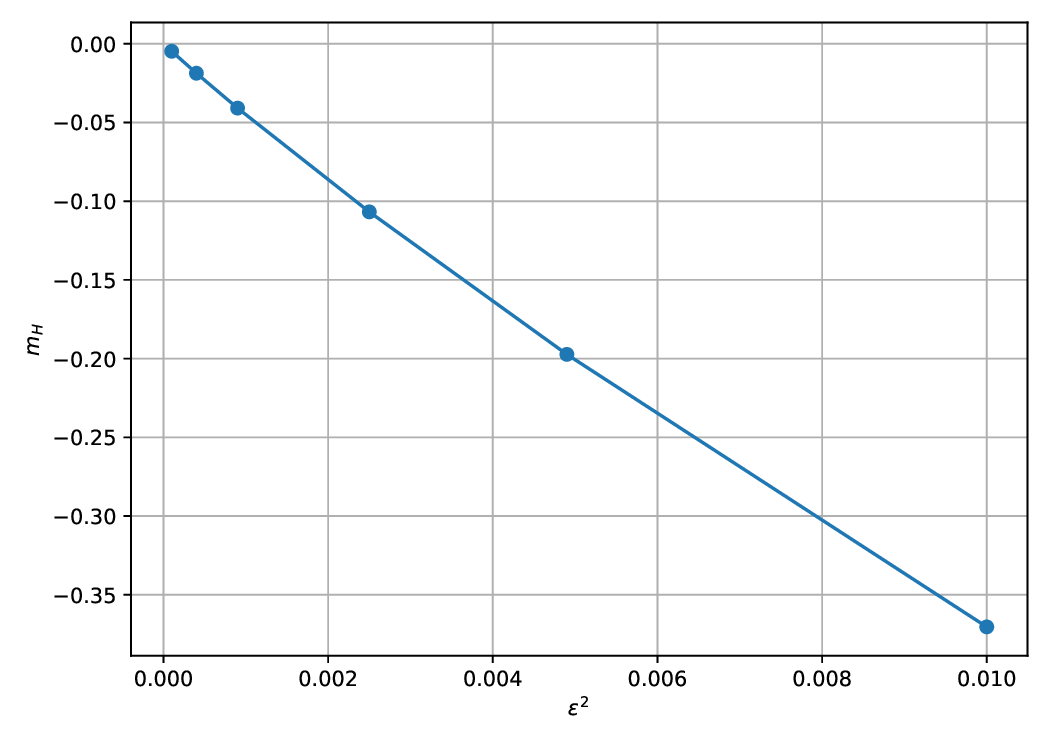}% 
\caption{\label{fig:epsart} Hawking mass $m_H$ of a perturbed sphere in Minkowski spacetime as a function of the perturbation amplitude squared $\epsilon^2$. The computed Hawking mass exhibits the expected approximately quadratic dependence on the perturbation amplitude for sufficiently small $\epsilon$, consistent with the perturbative expansion. }
\end{figure}

\begin{figure}
\centering
\includegraphics[width=100mm]{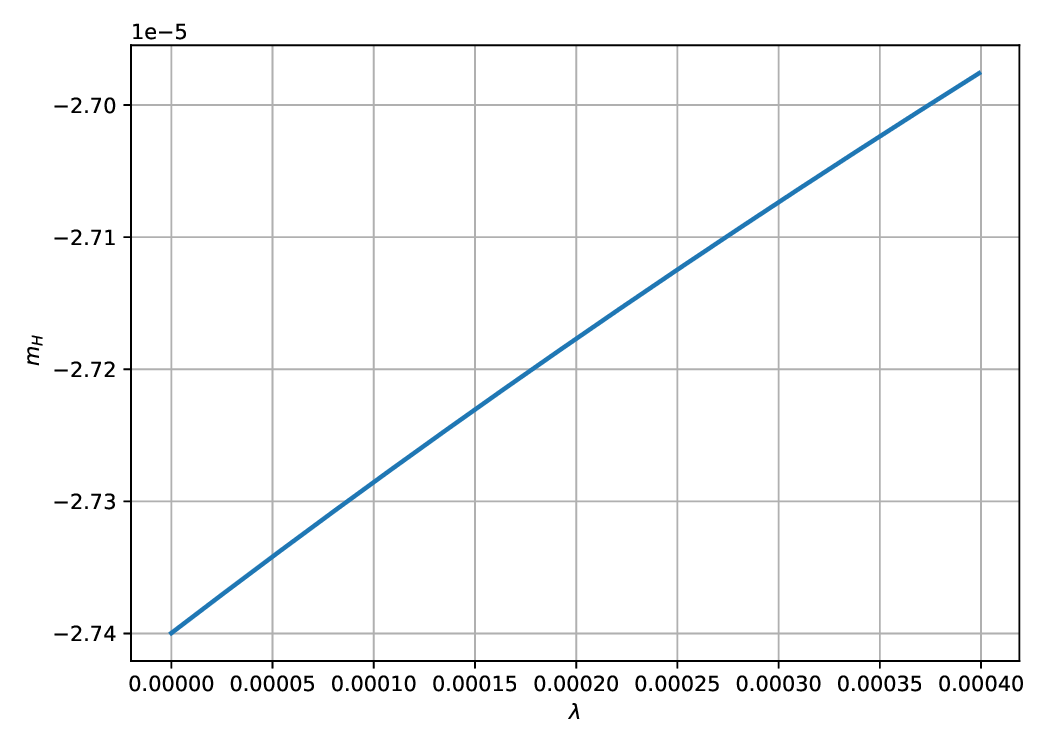}% 
\caption{\label{fig:epsart}Evolution of the Hawking mass for a perturbed sphere in Minkowski spacetime.  The Hawking mass increases monotonically throughout the evolution up to numerical tolerance, with no significant monotonicity violations or numerical instabilities observed over the evolution interval. }
\end{figure}

\subsection{Mesh refinement and convergence behaviour}

\noindent
To check the robustness of the numerical implementation, we performed a mesh refinement study for the flow described in the previous sections.  The evolution was repeated for a sequence of angular resolutions $(N_\theta,N_\phi)$ whilst keeping all physical and flow parameters fixed.  For each run, we monitored the quantity
\[
\min_n \left(m_H^{n+1}-m_H^n\right),
\]
corresponding to the largest decrease in Hawking mass observed during the evolution.  Fig. 4 shows the magnitude of this quantity as a function of angular resolution.  The monotonicity violations remain uniformly small across all tested meshes at the level of approximately between $10^{-9}$ and $10^{-8}$.  An initial decrease under refinement is observed, consistent with the interpretation that the residual violations arise primarily from discretization and stabilization effects, rather than any genuine physical effects or instabilities of the flow.  At higher resolutions are extremely small but do not decrease monotonically, which we attribute to competition between spatial discretization and the aritifical viscosity term.

\begin{figure}
\centering
\includegraphics[width=100mm]{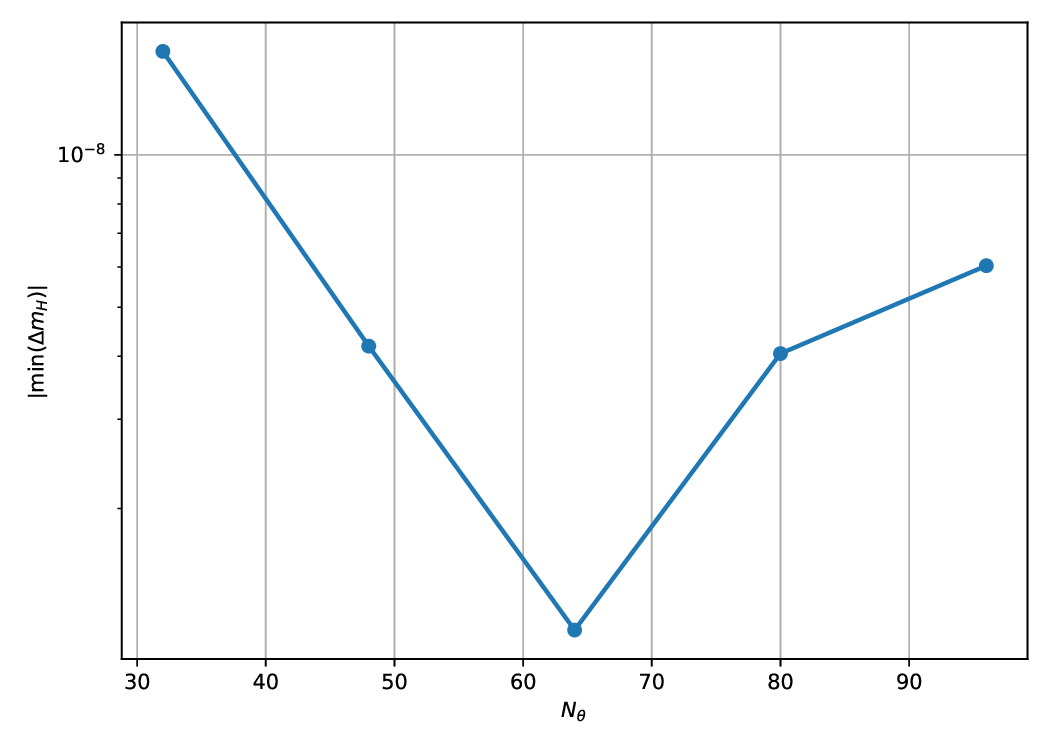}% 
\caption{\label{fig:epsart} Magnitude of the largest monotonicity violation $|\min_n(m_H^{n+1}-m_H^n)|$ during the stabilized spacetime evolution as a function of angular resolution.  The violations remain uniformly small across all tested meshes and decrease initially under refinement, indicating that the residual departures from exact monotonicity are dominated by numerical discretization effects. }
\end{figure}

\section{Spacetime perturbations}

\subsection{Perturbations of extrinsic curvature}

\noindent
To investigate departures from the totally geodesic setting, we introduce a perturbation of the hypersurface extrinsic curvature. Rather than solving the full Einstein constraint equations, we adopt a simplified perturbative model designed to probe the stability of Hawking mass monotonicity under small spacetime deformations.  The spacetime contribution is modeled via a scalar quantity $P$ representing the projection of the extrinsic curvature onto the evolving surface. In the numerical implementation, we take
\begin{equation}
P = \alpha \, (r - R_0),
\end{equation}
where $R_0$ denotes the radius of the unperturbed sphere and $\alpha$ is a small parameter controlling the strength of the spacetime perturbation.  This quantity modifies the magnitude of the spacetime mean curvature:
\begin{equation}
|\vec H|^2 = H^2 - P^2,
\end{equation}
where $H$ is the spatial mean curvature of the surface within the hypersurface. The case $\alpha=0$ corresponds to a totally geodesic hypersurface and recovers the IMCF in a Euclidean hypersurface considered previously. Nonzero values of $\alpha$ introduce a controlled departure from this scenario, allowing the flow to probe a regime which is genuinely spacetime-dependent whilst remaining numerically tractable.

Fig. 5 shows the evolution of the Hawking mass for a range of values for the spacetime perturbation parameter $\alpha$. When $\alpha = 0$, the evolution reduces to the time-symmetric case. As $\alpha$ increases, the curves undergo an upward shift whilst retaining monotone behaviour throughout the evolution interval considered.  Across all cases studied, no significant violations of Hawking mass monotonicity were observed beyond numerical tolerance. These results indicate that monotonicity of the Hawking mass persists under controlled departures from total geodesic hypersuraces and suggest that the underlying geometric mechanism is robust against small perturbations of the spacetime itself.

\begin{figure}
\centering
\includegraphics[width=100mm]{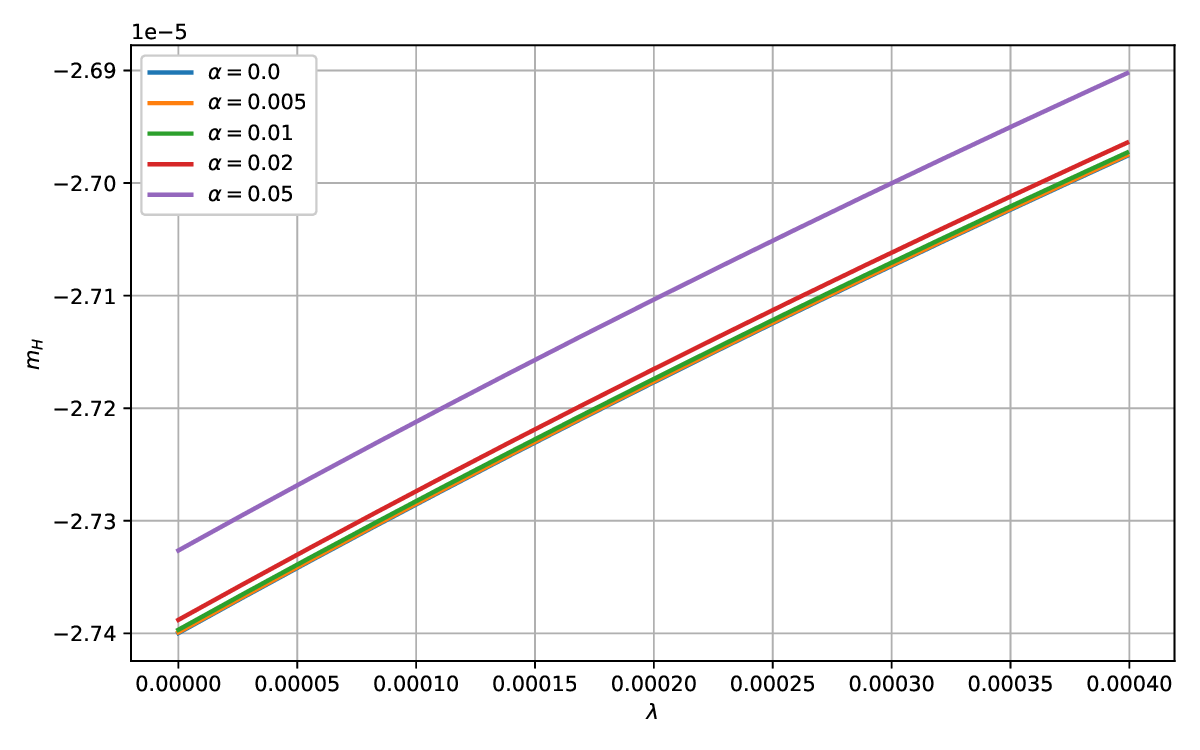}% 
\caption{\label{fig:epsart} Evolution of the Hawking mass under the spacetime-perturbed flow for several values of the extrinsic curvature parameter $\alpha$. The case $\alpha=0$ to the limit where the hypersurface-restricted IMCF is recovered. Increasing $\alpha$ introduces deviation via a nontrivial spacetime contribution to the mean curvature vector. In all cases shown, the Hawking mass remains monotone within numerical tolerance. }
\end{figure}

\subsection{Dependence on perturbation amplitude}

\noindent
We next wish to verify that the observed monotonicity is not an artifact of using a very small perturbation.  To do this, we repeated the spacetime flow evolution for a range of perturbation amplitudes $\epsilon$.  The initial surfaces were taken to be radial graph perturbations restricted to $Y_{20}$ as above and $\alpha$ was fixed in the below equation:
\begin{equation}
K_{ij}
=
\alpha\,Y_{20}(\theta,\phi)\,\delta_{ij}.
\end{equation}

\noindent
For a range of perturbation amplitudes up to $\epsilon = 0.1$, the numerical evolution remained stable and the Hawking mass was observed to evolve monotonically within numerical tolerance throughout the flow interval (shown in Fig. 6).  Increasing $\epsilon$ produced larger deviations of the initial surface from spherical symmetry and shifted the overall curve, but did not qualitatively change the monotonicity behaviour.  These results indicate that the observed monotonicity is robust across a finite perturbative regime and is not restricted solely to very small deformations.  For substantially larger perturbations, small monotonicity violations begin to appear, which we interpret as being due to breakdown of the radial graph approximation and stabilization scheme, rather than evidence against monotonicity for the underlying geometric flow itself.

\begin{figure}
\centering
\includegraphics[width=100mm]{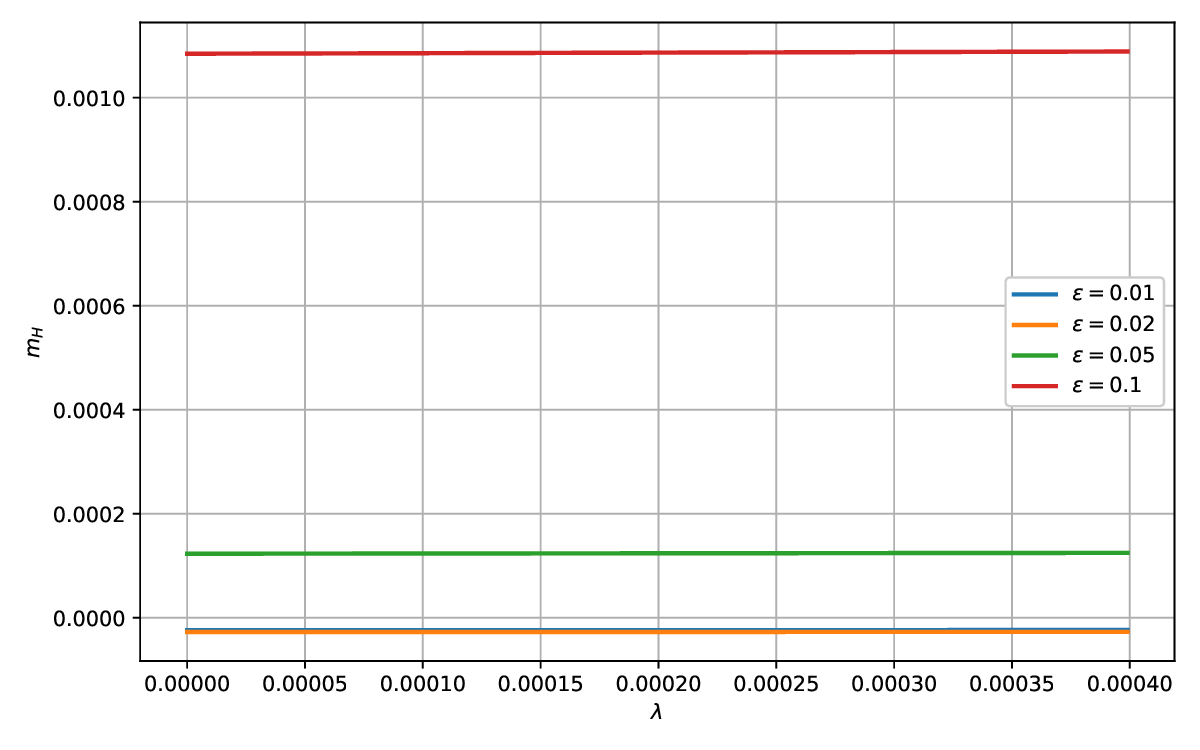}% 
\caption{\label{fig:epsart}Hawking mass $m_H(\lambda)$ during the spacetime UEF for several perturbation amplitudes $\epsilon$.  The initial surfaces are perturbed radial graphs of the form $r = R_0(1+\epsilon Y_{20})$ with fixed spacetime perturbation parameter $\alpha$.  Monotonicity is preserved within numerical tolerance across the perturbative regime considered.}
\end{figure}

\subsection{Dependence on angular perturbation sector}

\noindent
To investigate the dependence of the spacetime flow on angular structure, we repeated the numerical evolution for several spherical harmonic perturbations of the form
\begin{equation}
r(\theta,\phi)
=
R_0\left(1+\epsilon Y_{\ell 0}(\theta,\phi)\right),
\end{equation}
with fixed perturbation amplitude $\epsilon$ and fixed spacetime perturbation parameter $\alpha$.  The numerical results (shown in Fig. 7) demonstrate that  monotonicity is preserved up to numerical tolerance across multiple sectors, including the modes $Y_{10}$, $Y_{20}$, $Y_{30}$, and $Y_{40}$.  Higher harmonics produce larger angular gradients, leading to slightly larger deviations from exact monotonicity.  Nevertheless, the overall monotone behaviour of the Hawking mass remains stable and the observed monotonicity persists across a range of nonspherical perturbations.

\begin{figure}
\centering
\includegraphics[width=100mm]{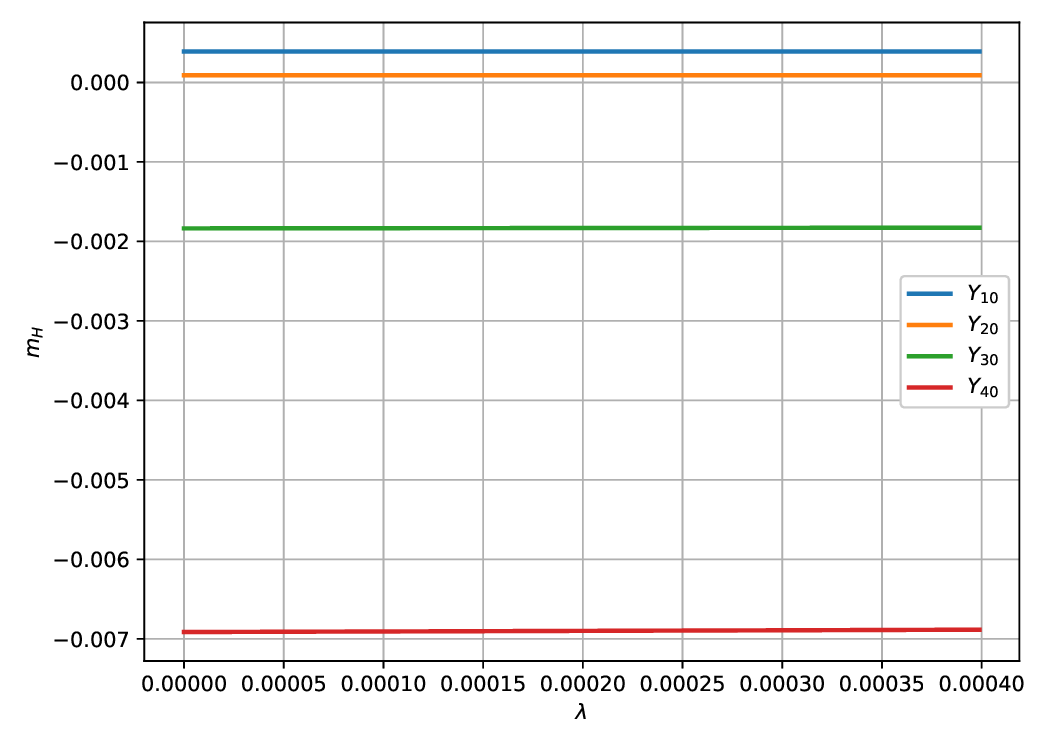}% 
\caption{\label{fig:epsart}Hawking mass $m_H(\lambda)$ during the spacetime UEF for several spherical harmonics $Y_{\ell 0}$.  The perturbation amplitude $\epsilon$ and spacetime perturbation parameter $\alpha$ are held fixed whilst $\ell$ is varied.  Monotonicity is preserved within numerical tolerance across all modes considered, although higher angular modes exhibit slightly larger deviations due to increased angular gradients and numerical stiffness.  }
\end{figure}

\section{Conclusions}

\noindent
In this work we developed and studied a perturbative numerical framework for UEFs in a spacetime setting, considering radial graph evolutions on perturbed  surfaces and examining the behaviour of the Hawking mass under small spacetime deformations.  A corrected scalar graph formulation of the flow suitable for the type of perturbation we considered was derived and implemented via an explicit numerical scheme.  This numerical framework was shown to recover expected results for round spheres in Minkowski space and the $\epsilon^2$ scaling of the Hawking mass.  The observed monotonicity violations remained uniformly small and were consistent with discretization effects.  We then introduced perturbative spacetime effects via a model which allows a small perturbative contribution to the extrinsic curvature.  This yielded a realization of a spacetime flow whilst remaining within a numerically tractable regime (a full discretization of the UEF equations is involved and beyond the scope of this work).  Across the range of perturbation amplitudes, spacetime deformation strengths and spherical harmonic modes considered here, the Hawking mass was observed to remain monotonic.

The results presented here should be interpreted as a proof-of-concept numerical study and we do not attempt a full nonlinear treatment of UEFs: the implementation is deliberately restricted to perturbative radial graph evolutions and short-time flows with stabilization.  Nevertheless, the calculations provide numerical evidence that monotonicity of the Hawking mass is preserved under small spacetime perturbations away from time-symmetric configurations.  There are various natural extensions to the work carried out here.  These include development of fully nonlinear spacetime implementations, discretization of the full UEFs equations in a numerically tractable form, investigations of long-time behaviour and singularity formation, and applications to more general asymptotically flat initial data sets.

\section*{Acknowledgments}

\noindent
HW thanks Béatrice Bonga for useful discussions.  This research was partly conducted whilst HW was visiting the Okinawa Institute of Science and 
Technology (OIST) through the Theoretical Sciences Visiting Program (TSVP).   HW acknowledges support from a London Mathematical Society Early Career Research Travel Grant (ECR-2526-51).      

\section*{Data availability}

\noindent
The numerical codes used to generate the results presented in this article are
available at \url{https://github.com/Tom467/HawkingmassUEFs}. No experimental data
were generated or analysed during the current study.

 \section*{Conflict of Interest}

\noindent
The authors have no conflicts to disclose.


\end{document}